\documentclass[twocolum,aps,pre,reprint,p]{revtex4-1}

\usepackage[dvips]{graphicx}
\usepackage{bm}
\usepackage{amsmath}
\usepackage{multirow}
\usepackage{color}
\usepackage{amssymb}

\begin{document}

\title{Interrelation between precisions on integrated currents and on recurrence times in Markov jump processes}

\thanks{
\textcopyright2025
American Physical Society. This is the accepted manuscript
of the following article: Alberto Garilli and Diego Frezzato, “Interrelation between precisions on integrated currents and on recurrence times in Markov jump processes,” {\em Phys. Rev. E},
Vol. 112, No. 4, 044141 (2025) The final published version is
available from DOI: https://doi.org/10.1103/27gn-7w5d.
}

\author{Alberto Garilli and Diego Frezzato{$^\ddagger$}}

\affiliation{Department of Chemical Sciences, University of Padova, via Marzolo 1, I-35131, Padova, Italy. $^\ddagger$Email: diego.frezzato@unipd.it}

%\email{diego.frezzato@unipd.it}

\date{\today}

%%%%%%%%%%%%%%%%%%%%%%%%%%%%%%%%%%%%%%%%%%%%%%%%%%%%%%%%%%%%%%%%%%%

\begin{abstract} 
For Markov jump processes on irreducible networks with finite number of sites, we derive a general and explicit expression of the squared coefficient of variation for the net number of transitions from one site to a connected site in a given time window of observation (i.e., an `integrated current' as dynamical output). Such expression, which in itself is particularly useful for numerical calculations, is then elaborated to obtain the interrelation with the precision on the intrinsic timing of the recurrences of the forward and backward transitions. 
In biochemical ambits, such as enzyme catalysis and molecular motors, the precision on the timing is quantified by the so-called randomness parameter and the above connection is established in the long time limit of monitoring and for an irreversible site-site transition; the present extension to finite time and reversibility adds a new dimension. Some kinetic and thermodynamic inequalities are also derived.      
\end{abstract}

\maketitle

\section{Context and motivation}

A wealth of dynamical processes in various ambits of natural sciences can be effectively modeled as continuous-time Markov jump processes among a finite number $N$ of sites. For instance, in chemical contexts, such a model is able to grasp the slow transitions among conformational energy wells \cite{Moro1989, Loutchko16}, the jumps of tagged molecular moieties among hosting species \cite{skodje1, Angew2019, MB2021, JCP2022}, the transitions in the copy-number space for reactive systems involving low numbers of molecules \cite{gillespie2007}, to describe hopping processes \cite{Derrida1983}, the operation of molecular motors \cite{Block95, Fisher07, Kolo2000}, features of complex biochemical networks \cite{Banerjee17, Mallory19, Mallory2020}, and more. In the simplest and most relevant setup, to which we shall adhere in this work, the jump rate constants from site to site are time-independent and the network is irreducible, i.e., there is at least one path to go from one site to any other one; in this situations, the process admits a unique stationary distribution with occupation probabilities $p_i^{\rm ss} > 0$.

Let us introduce the oversimplified notation that will be used throughout. To be general, we shall admit that the site-to-site jumps can occur via multiple transition channels, and assume to be able to distinguish such channels within a given degree of resolution (in fact, a channel may be a real physical way of jump, or may result from the lumping of unresolved channels). 
For instance, this can be important in (bio)chemical ambits where a jump from a molecular state to another can be due to different reactions.
Talking of a transition, say $\alpha \to \beta$, it will be implicit that we are referring to one of the channels (possibly only one) to go from $\alpha$ to $\beta$.
In particular, $k$ will stand for rate constants of specific channels, while $k^{\rm tot}$, where needed, will denote the total jump rate constant (sum over the channels).

\begin{figure}
	\centering
	\includegraphics[width = 0.95\linewidth]{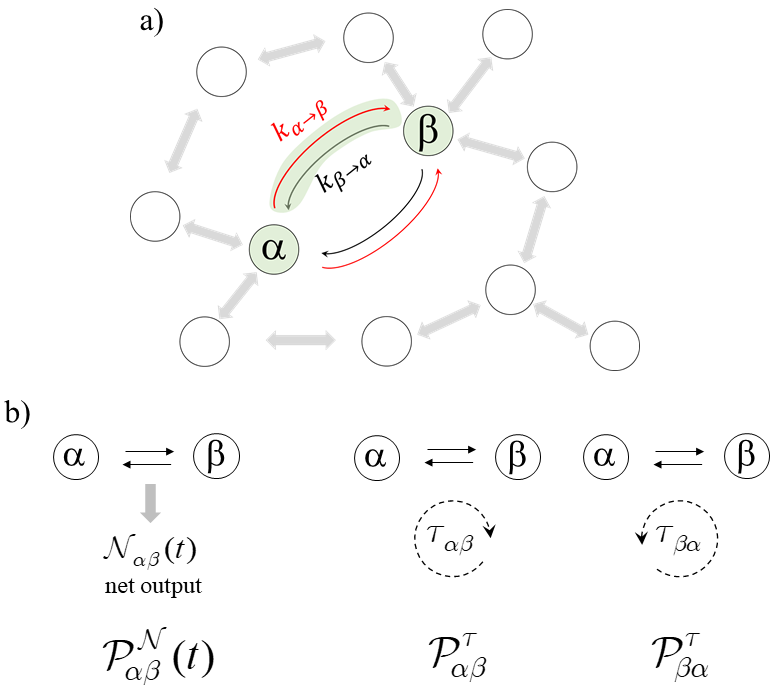} 
\caption{a) Pictorial representation of the $\alpha \leftrightarrow \beta$ jumps; $k_{\alpha \to \beta}$ and $k_{\beta \to \alpha}$ are the jump rate constants for the specific channel under consideration. b) The precision coefficients concerning the $\alpha \leftrightarrow \beta$ channel. The circular dashed arrows stand for the repetition of the transitions $\alpha \to \beta$ or $\beta \to \alpha$; note that before a transition is repeated, the backward one (if feasible) could occur several times.}
	\label{Fig1}
\end{figure}

An ambit that has attracted attention in recent years is the characterization of steady-state integrated currents, i.e., net outputs in a given time-window of observation with no information about the past history of the system. Let us introduce the specific dynamical output of interest here. Imagine monitoring the forward/backward transitions between a pair of sites $\alpha$ and $\beta$ directly connected by a transition channel. Figure \ref{Fig1} gives a pictorial representation. To be general we assume both $\alpha \to \beta$ and $\beta \to \alpha$, but the bidirectionality is not mandatory except when explicitly stated. Let ${\cal N}_{\alpha \beta}(t)$ be the {\em net} number of jumps from $\alpha$ to $\beta$ in an observation time $t$ (as said above, the starting condition is meant to be sampled at stationarity). Such a number is a stochastic variable with a statistical distribution having $t$-dependent moments $\langle {\cal N}_{\alpha \beta}(t)^n \rangle$. The average is simply
\begin{eqnarray}
\langle {\cal N}_{\alpha \beta}(t)\rangle = J_{\alpha\beta} \, t    
\end{eqnarray}
with $J_{\alpha\beta}$ the steady-state probability current in the direction $\alpha$-to-$\beta$, i.e.,
\begin{equation}
J_{\alpha\beta} := F_{\alpha\beta} - F_{\beta\alpha}    
\end{equation}
where $F_{ij} = p_i^{\rm ss} k_{i \to j}$ is the steady-state probability flux from $i$ to $j$ over the specific transition channel. While $\langle {\cal N}_{\alpha \beta}(t)\rangle$ for given $t$ has to do with the steady-state {\em speed} $J_{\alpha\beta}$ of the output's production, the following ratio, built with average and variance, is typically used to quantify the {\em precision} on the output:
\begin{equation}\label{eq_CVdef}
{\cal P}_{\alpha\beta}^{\cal N}(t) := \frac{\langle {\cal N}_{\alpha \beta}(t)^2 \rangle - \langle {\cal N}_{\alpha \beta}(t)\rangle^2}
{\langle {\cal N}_{\alpha \beta}(t)\rangle^2}    
\end{equation}
Ratios of such a type, known as squared coefficients of variation, will be here termed `precision coefficients' and denoted by the letter ${\cal P}$.
Lower bounds on ${\cal P}_{\alpha\beta}^{\cal N}(t)$ of kinetic \cite{Baiesi2019}, thermodynamic \cite{TUR1, TUR2, TUR3} and kinetic-thermodynamic \cite{Vo2022, Frezzato2020} type have been derived in the past years. In particular, the kinetic uncertainty relation \cite{Baiesi2019} states that  ${\cal P}_{\alpha\beta}^{\cal N}(t) \ge (\kappa t)^{-1}$ where $\kappa = \sum_{i, j \neq i} p_i^{\rm ss} k_{i \to j}^{\rm tot}$ is the global activity of the network (average number of jumps per unit of time). In a network where all channels are bidirectional, the thermodynamic uncertainty relation (TUR) \cite{TUR1, TUR2} states instead that ${\cal P}_{\alpha\beta}^{\cal N}(t) \ge 2 (\sigma^{\rm ss} t)^{-1}$ where $\sigma^{\rm ss}$ is the steady-state average rate of entropy production (in units of Boltzmann constant) given by Schnakenberg's expression \cite{net1976} taking into account all channels that we are able to discern \cite{note1}.

% ----------------------- 
% FIGURE 1
% ----------------------- 

In parallel, we may consider the statistics of the recurrence time of the site-to-site transitions, i.e., the time waited before a given transition occurs again. Note that before a given transition takes place again, the backward transition (if feasible) could occur several times. Let $\tau_{\alpha\beta}$ and  $\tau_{\beta\alpha}$ be the recurrence times of $\alpha \to \beta$ and $\beta \to \alpha$, respectively. Such times are aleatory variables statistically distributed with moments $\overline{\tau_{\alpha\beta}^n}$ and $\overline{\tau_{\beta\alpha}^n}$ (for this kind of averages we prefer to use the overbar in place of angular brackets \cite{note_overbar}). In particular, the average values are directly related to the steady-state probability fluxes via \cite{Frezzato2020}
\begin{equation}\label{taus}
\overline{\tau}_{\alpha\beta} = F_{\alpha\beta}^{-1}  \;\; , \;\;
\overline{\tau}_{\beta\alpha} = F_{\beta\alpha}^{-1} 
\end{equation}
With averages and second moments we can build the following precision coefficients on the timing of the two recurrences:
\begin{eqnarray}\label{CVtau}
{\cal P}_{\alpha \beta}^\tau := \frac
{\overline{\tau^2_{\alpha\beta}} - \overline{\tau}_{\alpha\beta}^2}
{\overline{\tau}_{\alpha\beta}^2}  \;\; , \;\;
{\cal P}_{\beta\alpha}^\tau := \frac
{\overline{\tau^2_{\beta\alpha}} - \overline{\tau}_{\beta\alpha}^2}
{\overline{\tau}_{\beta\alpha}^2}
\end{eqnarray}

It is evident that there must be an interrelation between ${\cal P}_{\alpha\beta}^{\cal N}(t)$ on one side and ${\cal P}_{\alpha\beta}^\tau$ and ${\cal P}_{\beta\alpha}^\tau$ on the other side, although the two types of coefficients have quite different properties. 
In fact, ${\cal P}_{\alpha\beta}^{\cal N}(t)$ concerns the extensive net output and depends on time, whereas the coefficients ${\cal P}^\tau$ refer to the intrinsic recurrence times. Moreover, ${\cal P}_{\alpha\beta}^{\cal N}(t)$ is defined only for out-of-equilibrium steady states for which the average current is non-null, whereas ${\cal P}^\tau$ is also defined at equilibrium (with ${\cal P}_{\alpha\beta}^\tau={\cal P}_{\beta\alpha}^\tau$ \cite{JCP2019}). A remarkable theorem \cite{Erlang} states that, in a network with $N$ states, ${\cal P}^\tau \ge N^{-1}$ for any monitored transition channel regardless of the topology of the network (the equality holds in the unicyclical case with irreversible transitions). This surprising bound sets an intrinsic and general relationship between the precision on any transition's recurrence and the size of the network.

Both types of precision coefficients are known in the field of statistical chemical kinetics \cite{Moffitt14, Moffitt10a} whose main goal is making inferences about the underlying reaction mechanism having a few experimental observations at disposal. For instance, widely studied cases are the enzyme catalysis (where $\alpha \leftrightarrow \beta$ corresponds to the reaction channel of product's formation) and the operation of processive molecular motors ($\alpha \leftrightarrow \beta$ corresponds to translational or rotational steps). In particular, the bound on ${\cal P}_{\alpha\beta}^\tau$ mentioned above is useful to establish the minimal number of states that {\em must} be present in the underlying mechanisms \cite{Moffitt14, Moffitt10b, Fisher07}. 
A crucial point is how to experimentally assess the precision coefficients ${\cal P}^\tau$. Although single-molecule techniques nowadays allow to monitor the operation of systems such as rotary motors \cite{REF_ROT1, REF_ROT2} and intracellular transporters \cite{REF_KIN1, REF_KIN2} on the timeline, and hence to have direct access to the statistics of the recurrences, one would desire a connection of the ${\cal P}^\tau$ with the statistics of the extensive ${\cal N}_{\alpha\beta}(t)$ at given $t$. In this regard it has been shown \cite{Block95} that in the limit of infinitely long observation time and for networks in which $\alpha \to \beta$ (the cycle's completion step) is irreversible, the coefficient ${\cal P}_{\alpha\beta}^\tau$, also known as `randomness parameter', is experimentally achievable by exploiting its equivalence with the Fano factor \cite{Moffitt14} which corresponds to $\langle {\cal N}_{\alpha \beta}(t)\rangle \times {\cal P}_{\alpha\beta}^{\cal N}(t)$ as $t \to \infty$. 
On the other hand, while the crucial $\alpha \leftrightarrow \beta$ in enzymes and processive motors is practically unidirectional under normal conditions, there might be situations in which backward steps can occasionally take place and in principle cannot be ignored; for instance, backward steps have been seen in the rotary $\rm F_1$-ATPase motor with an attached actin filament at low ATP concentration \cite{REF_F1motor_back}, and in kinesins under sufficiently high opposing loads \cite{REF_kine_back}. In summary, bidirectionality must be taken into account if the backward steps cannot be kinetically neglected. In addition, the finiteness of the observation time might introduce an extra potentially useful dimension.

Besides the biological contexts mentioned above just as examples, we stress that the problem of connecting the two types of precision coefficients regards any Markov jump process in irreducible networks of finite dimension with fixed rate constants. Here we obtain such desired connection (see Eq. \ref{eq_corr} later), valid for finite times and generic networks, in which ${\cal P}_{\alpha\beta}^{\cal N}(t)$ is related to the two ${\cal P}_{\alpha\beta}^\tau$ and ${\cal P}_{\beta\alpha}^\tau$ for bidirectional transitions, or to the single ${\cal P}_{\alpha\beta}^\tau$ for one-directional $\alpha \to \beta$. Apart from retrieving the known result for irreversible $\alpha \to \beta$ and $t \to \infty$ as a special case, the general relation seems to be a promising starting point for deriving interrelations and mutual bounds between (thermo)dynamical features of the network. In this regard, some preliminary kinetic and thermodynamic bounds (the latter obtained by exploiting the TUR) will be presented and illustrated for a simple 4-site network.

In addition to the specific practical target outlined above, this work also bears a methodological relevance concerning the derivation of an expression of ${\cal P}_{\alpha\beta}^{\cal N}(t)$ (see Eq. \ref{eq_CV} later) which makes use of the moment generating function method \cite{Polettini2019} as shown in Appendix \ref{AppA}. Such expression allows to easily get the limits of ${\cal P}_{\alpha\beta}^{\cal N}(t)$ as $t \to 0$ and $t \to \infty$, and is particularly suitable for numerical calculations at intermediate times where explicit analytical forms cannot be achieved.

\section{Results}\label{sec_results}

\subsection{Preliminaries}

Let us introduce some quantities that will appear later. Let $\epsilon$ be the `rectifying efficiency' of the $\alpha \leftrightarrow\beta$ transition channel defined as
\begin{equation}\label{eq_eps}
\epsilon := \frac{F_{\alpha\beta} - F_{\beta\alpha}}{F_{\alpha\beta} + F_{\beta\alpha}}    
\end{equation}
The numerator is the average probability current $J_{\alpha\beta}$ while the denominator gives the average number of jumps $\alpha \leftrightarrow \beta$ per unit of time (i.e., the activity on the transition channel under consideration). Note that $-1 \leq \epsilon \leq +1$ where the extrema $+1$ and $-1$ correspond, respectively, to the one-directional situations $\alpha \to \beta$ and $\beta \to \alpha$, while $\epsilon=0$ if the forward and backward fluxes are equal.

Then, let us introduce the following time-dependent quantifier of the relative deviation from the stationary distribution conditioned by the knowledge about the system's state at a previous time-zero (in what follows, $t$ is the temporal separation from such initial instant):
\begin{equation}\label{eq_chi}
\chi_{i s_0}(t) := \frac{ p(i,t|s_0) - p_i^{\rm ss}}{p_i^{\rm ss}}    
\end{equation}
where $ p(i,t|s_0)$ is the probability of being in the site $i$ at the time $t$ if the system was in $s_0$ at the time-zero. The initial condition is
$\chi_{i s_0}(0) = (\delta_{i,s_0} - p_i^{\rm ss})/p_i^{\rm ss}$ with $\delta$ the Kronecker's delta, while $\lim_{t \to \infty} \chi_{i s_0}(t) = 0$.

In Appendix \ref{AppB} it is shown that
\begin{eqnarray}\label{eq_rec}
\int_0^t dt' \chi_{i s_0}(t') = -\overline{\tau}_{ij|s_0} 
+ \sum_n \overline{\tau}_{ij|n} \, p(n,t|s_0)
\end{eqnarray}
where $\overline{\tau}_{ij|s_0}$ is the average {\em occurrence} time of the $i \to j$ transition starting from the generic site $s_0$; taking $s_0 = j$ we have $\overline{\tau}_{ij|j} \equiv \overline{\tau}_{ij}$, i.e., the average {\rm recurrence} time already introduced (see Eq. \ref{taus}). The integral in Eq. \ref{eq_rec} will play a crucial role later, and can be further elaborated. From Eq. \ref{eq_chi} we get $p(n,t|s_0) = p_n^{\rm ss} (\chi_{n s_0}(t) +1)$, which, when plugged into Eq. \ref{eq_rec}, leads to
\begin{eqnarray}\label{eq_rec2}
\int_0^t dt' \chi_{i s_0}(t') &=& -\overline{\tau}_{ij|s_0} 
+ \frac{\overline{\tau}_{ij}}{2} \left( 1 + {\cal P}_{ij}^\tau \right) \cr
&+& \sum_n \overline{\tau}_{ij|n} \, p_n^{\rm ss} \, \chi_{n s_0}(t)
\end{eqnarray}
where it has been made use of the property \cite{Frezzato2020}
\begin{equation}
\sum_n \overline{\tau}_{ij|n} \, p_n^{\rm ss} = \frac{\overline{\tau^2_{ij}}}{{2 \overline{\tau}_{ij}}}    
\end{equation}
and of the definition ${\cal P}_{ij}^\tau = ({\overline{\tau^2_{ij}}}- \overline{\tau}_{ij}^2)/\overline{\tau}_{ij}^2$.  
Taking $t \to \infty$, the integral converges to $-\overline{\tau}_{ij|s_0} 
+ \overline{\tau}_{ij} ( 1 + {\cal P}_{ij}^\tau) /2$; this will be useful to determine the asymptotics of the precision coefficients in the long-time limit.

Remarkably, Eqs. \ref{eq_rec} and \ref{eq_rec2} hold for any choice of site $j \neq i$ directly reachable from $i$, and for any transition channel connecting $i$ to $j$ (if there are multiple channels). This gives us freedom to make the most appropriate choice depending of the specific use of Eq. \ref{eq_rec2}.

For any pair of sites $i$ and $j \neq i$ directly reachable from $i$, the following bounds hold:
\begin{eqnarray}
&&-(\overline{\tau}_{ij} - \overline{\tau}_{ij}^{\rm min}) \leq 
\int_0^t dt' \chi_{i j}(t') \leq 
\overline{\tau}_{ij}^{\rm max}  - \overline{\tau}_{ij} \; , \label{eq_bxi1} \\
&&0 \leq  \int_0^t dt' \chi_{i i}(t') \leq 
\overline{\tau}_{ij}^{\rm max}  - \overline{\tau}_{ij}^{\rm min} \label{eq_bxi2}
\end{eqnarray}
where
\begin{equation}\label{eq_taumax}
\overline{\tau}_{ij}^{\rm max} := \max_{n} \{ \overline{\tau}_{ij|n} \} \;\; , \;\;
\overline{\tau}_{ij}^{\rm min} := \min_{n} \{ \overline{\tau}_{ij|n} \} = \overline{\tau}_{ij|i}
\end{equation}
These bounds follow directly from Eq. \ref{eq_rec} with $s_0 = j$ (for Eq. \ref{eq_bxi1}) or $s_0 = i$ (for Eq. \ref{eq_bxi2}). We have $\overline{\tau}_{ij}^{\rm min} = \overline{\tau}_{ij|i}$ because the $i \to j$ occurrence is on average surely faster starting already from $i$ than starting from any other site.

\subsection{Precision coefficient for the integrated current}\label{subsec_p1}

The following expression is derived in Appendix \ref{AppA}: 
\begin{equation}\label{eq_CV}
{\cal P}_{\alpha\beta}^{\cal N}(t) = \frac{1}{\epsilon J_{\alpha\beta} t} 
- \frac{1}{t^2}\int_0^t dt' \int_0^{t'} dt'' \, \gamma (t'')    
\end{equation}
where 
\begin{eqnarray}\label{eq_gamma}
\gamma(t) = 
c_0 \chi_{\alpha\alpha}(t) 
- c_+ \chi_{\alpha\beta}(t)
- c_- \chi_{\beta\alpha}(t)
+ c_0 \chi_{\beta\beta}(t) 
\cr
\end{eqnarray}
with the $\chi$'s defined in Eq. \ref{eq_chi}, and with the following non-negative dimensionless coefficients related to the rectifying efficiency:
\begin{eqnarray}\label{eq_cs}
c_0 =  \frac{1-\epsilon^2}{2 \epsilon^2} \;\; , \;\;
c_{\pm} = \frac{(1 \pm \epsilon)^2}{2 \epsilon^2}  
\end{eqnarray}
Note that $c_0 \to 0$ as $|\epsilon| \to 1$, while $c_\pm \to 2$ as  $\epsilon \to \pm 1$, and $c_\pm \to 0$ as $\epsilon \to \mp 1$. 

The short-time limit is readily obtained from Eq. \ref{eq_CV}:
\begin{equation}\label{eq_CV_shortt}
{\rm short} \; t: \;\; {\cal P}_{\alpha\beta}^{\cal N}(t) \simeq \frac{1}{\epsilon J_{\alpha\beta} t} + r   
\end{equation}
where $r = -\gamma(0)/2 <0$ because $\gamma(0) >0$ \cite{note_gamma0}. As $t \to 0$, the offset $r$ becomes negligible and the precision coefficient grows hyperbolically. As show later, we have that ${\cal P}_{\alpha\beta}^{\cal N}(t) \propto t^{-1}$ also at long $t$, but with a proportionality coefficient lower or higher than $(\epsilon J_{\alpha\beta})^{-1}$. A more complex behavior is expected in the intermediate time window.

Let us stress that the terms of the kind $\chi_{is_0}(t)$ that enter $\gamma(t)$ in Eq. \ref{eq_gamma} are directly related with the steady-state occupation probabilities $p_i^{\rm ss}$ and with the time-dependent conditional probabilities $p(i,t|s_0)$ through Eq. \ref{eq_chi}. Both these quantities are easily obtained by means of consolidated computational methods, making Eq. \ref{eq_CV} particularly suited for numerical calculations of ${\cal P}_{\alpha\beta}^{\cal N}(t)$, especially in the intermediate timescale where analytical solutions cannot be achieved. In particular, in Appendix \ref{AppC} we outline the numerical route based on the spectral decomposition \cite{note3b} valid in the case of diagonalizable rate matrices (which is, in fact, the most typical situation).

\subsection{Interrelation between ${\cal P}_{\alpha\beta}^{\cal N}(t)$ and ${\cal P}_{\alpha\beta}^\tau$, ${\cal P}_{\beta\alpha}^\tau$}\label{subsec_p2}

Starting from Eq. \ref{eq_CV} with Eq. \ref{eq_gamma}, and making use of Eq. \ref{eq_rec2}, in Appendix \ref{AppD} it is shown that ${\cal P}_{\alpha\beta}^{\cal N}(t)$ can be cast in the form
\begin{eqnarray}\label{eq_corr}
{\cal P}_{\alpha\beta}^{\cal N}(t) = \frac{{\cal T}_\infty}{t} + 
\frac{1}{t^2}\int_0^t dt' \varphi(t')
\end{eqnarray}
where the characteristic time ${\cal T}_\infty = \lim_{t \, \to \infty} [t {\cal P}_{\alpha\beta}^{\cal N}(t)]$ and the characteristic function $\varphi(t)$ (which also has physical dimension of time) take on different forms depending on whether $\alpha$ and $\beta$ are connected by a bidirectional transition channel, or the transition from $\alpha$ to $\beta$ is one-directional. In the bidirectional case we have 
\begin{eqnarray}\label{tauinft_bid}
{\cal T}_\infty = -\frac{1}{\epsilon J_{\alpha\beta}} + 
\frac{{\cal P}_{\alpha\beta}^\tau  - {\cal P}_{\beta\alpha}^\tau}{J_{\alpha\beta}}
+ c_0
\left( \overline{\tau}_{\alpha\beta|\alpha} + \overline{\tau}_{\beta\alpha|\beta}\right)
\end{eqnarray}
and
\begin{eqnarray}\label{eq_phi_bid}
\varphi(t)= c_0 \sum_n \left( 
\frac{\overline{\tau}_{\alpha\beta|n}}{\overline{\tau}_{\alpha\beta}} 
- \frac{\overline{\tau}_{\beta\alpha|n}}{\overline{\tau}_{\beta\alpha}}
\right) \times \cr   
\times \left( 
\overline{\tau}_{\beta\alpha} \chi_{n \beta}(t) -
\overline{\tau}_{\alpha\beta} \chi_{n \alpha}(t)  
\right) p_n^{\rm ss}
\end{eqnarray}
In the one-directional case we have instead
\begin{eqnarray}\label{tauinft_oned}
{\cal T}_\infty = {\cal P}_{\alpha\beta}^\tau \overline{\tau}_{\alpha\beta}
\end{eqnarray}
and
\begin{eqnarray}\label{eq_phi_oned}
\varphi(t)= 2 \sum_n \overline{\tau}_{\alpha\beta|n} \,
\chi_{n \beta}(t) \, p_n^{\rm ss}
\end{eqnarray}
It is worth noting that Eq. \ref{eq_corr} with Eqs. \ref{tauinft_oned}-\ref{eq_phi_oned} holds also if $\alpha \leftrightarrow \beta$ is bidirectional but we observe only $\alpha \to \beta$, i.e., if we take ${\cal N}_{\alpha\beta}$ to be the number of jumps from $\alpha$ to $\beta$ (not the net number of jumps). To see this, it suffices to repeat the derivation in Appendix \ref{AppA} by considering only the `counting field' $+q$ in Eq. \ref{eq_Ddef}.

Equations \ref{eq_corr}-\ref{eq_phi_oned} provide the interrelation between the two types of precision coefficients. Remarkably, the first addend ${\cal T}_\infty/t$ in Eq. \ref{eq_corr} contains only {\em local} dynamical observables (however implicitly dependent on the global dynamics) of the $\alpha \leftrightarrow \beta$ channel, while the second addend explicitly contains features of the rest of the network.

At long times, the second addend in Eq. \ref{eq_corr} is proportional to $t^{-2}$ because the integrals of the type in Eq. \ref{eq_rec2} converge to finite values and so also the integral of $\varphi(t')$ does. Hence,
\begin{equation}\label{eq_CV_longt}
{\rm long} \; t : \;\; {\cal P}_{\alpha\beta}^{\cal N}(t) \simeq 
\frac{{\cal T}_\infty}{t}     
\end{equation}
Equation \ref{eq_CV_longt} is the counterpart of Eq. \ref{eq_CV_shortt} in the long-time limit.

In the long-time limit of monitoring and for irreversible $\alpha \to \beta$, Eq. \ref{eq_CV_longt} allows us to retrieve the known relation \cite{Block95} between ${\cal P}_{\alpha\beta}^\tau$, seen as `randomness parameter', and the experimentally achievable quantity 
\begin{eqnarray}
r_{\alpha\beta} &=& \lim_{t \to \infty }{[\langle {\cal N}_{\alpha \beta}(t)^2 \rangle - \langle {\cal N}_{\alpha \beta}(t)\rangle^2}]/ {\langle {\cal N}_{\alpha \beta}(t)\rangle} \cr 
&\equiv& J_{\alpha\beta} \lim_{t \to \infty} [{\cal P}_{\alpha\beta}^{\cal N}(t) \times t]     
\end{eqnarray}
known as Fano factor. By using Eq. \ref{tauinft_oned} and $J_{\alpha\beta} \equiv F_{\alpha\beta} = \overline{\tau}_{\alpha\beta}^{-1}$, we obtain the known result 
\begin{equation}
r_{\alpha\beta} \equiv {\cal P}_{\alpha\beta}^\tau     
\end{equation}
However, Eq. \ref{eq_corr} with Eqs. \ref{tauinft_bid}  and \ref{eq_phi_bid} extends the interrelation between the different types of precision coefficients to finite times and reversibility.

\subsection{Kinetic and thermodynamic bounds}\label{subsec_p4}

By employing Eqs. \ref{eq_bxi1} and \ref{eq_bxi2} we can get a lower bound on the inner integral $\int_0^{t'} dt'' \gamma(t'')$ that enters Eq. \ref{eq_CV}. With a few steps (see ref. \cite{note_bound}), this leads to the following non-trivial kinetic {\em upper} bound:
\begin{eqnarray}\label{eq_kinbound}
{\cal P}_{\alpha\beta}^{\cal N}(t) \leq \frac{{\cal T}^{\rm ub}}{t}    
\end{eqnarray}
where
\begin{eqnarray}\label{eq_b0}
{\cal T}^{\rm ub} = -\frac{1}{\epsilon J_{\alpha\beta}}
+ c_+ \overline{\tau}^{\rm max}_{\alpha\beta} 
+ c_- \overline{\tau}^{\rm max}_{\beta\alpha} 
\end{eqnarray}
with the definition in Eq. \ref{eq_taumax}. The relation holds also for irreversible transitions \cite{note_bound} (for instance, by setting $\epsilon = 1$, $c_+ = 2$ and $c_- =0$ in the case of only $\alpha \to \beta$). Note that Eq. \ref{eq_kinbound} can be further elaborated obtaining the weaker bound with ${\cal T}^{\rm ub}$ replaced by ${\cal T}^{{\rm ub} '} = -(\epsilon J_{\alpha\beta})^{-1} + (1+\epsilon^2) \epsilon^{-2} F_{\rm min}^{-1}$ in which $F_{\rm min}$ is the lowest steady-state probability flux in the network. 

A kinetic inequality involving ${\cal P}_{\alpha\beta}^\tau  - {\cal P}_{\beta\alpha}^\tau$ can be obtained from Eq. \ref{tauinft_bid} by enforcing ${\cal T}_\infty \geq 0$ and using $\overline{\tau}_{\alpha\beta|\alpha} < \overline{\tau}_{\alpha\beta}$ and $\overline{\tau}_{\beta\alpha|\beta} < \overline{\tau}_{\beta\alpha}$. A few algebraic steps lead to the relation
\begin{eqnarray}\label{eq_b4}
\epsilon \, ({\cal P}_{\alpha\beta}^\tau  - {\cal P}_{\beta\alpha}^\tau)
\; > - 1   
\end{eqnarray}
which is non-trivial since the quantity on the left-hand side can be either positive or negative. 

Thermodynamic inequalities can be obtained by exploiting the TUR in the specific case of networks with site-site connections all bidirectional. Since Eqs. \ref{eq_CV_shortt} and \ref{eq_CV_longt} become exact, respectively, for $t\to 0$ and $t \to \infty$, from the TUR we get $\epsilon J_{\alpha\beta} \leq \sigma^{\rm ss}/2$ and ${\cal T}_{\infty} \geq 2/\sigma^{\rm ss}$. 
The short-time bound gives $|J_{\alpha\beta}| \leq \sqrt{\sigma^{\rm ss} b/2}$
where $b = F_{\alpha\beta} + F_{\beta\alpha}$ is the dynamical activity over the $\alpha \leftrightarrow \beta$ channel.  
Inequalities of such a type appear in regard of the average speed of processive motors \cite{Piet2016, Li2020}, but with reference to the opposite $t \to \infty$ limit. 
The elaboration of the lower bound on ${\cal T}_{\infty}$ leads instead to the following inequality which contains the weaker Eq. \ref{eq_b4} \cite{note5}: 
\begin{eqnarray}\label{eq_b3}
\epsilon \, ({\cal P}_{\alpha\beta}^\tau  - {\cal P}_{\beta\alpha}^\tau)
\; > \;
\frac{2 \epsilon J_{\alpha\beta}}{\sigma^{\rm ss}} - 1   
\end{eqnarray}

A numerical exploration of the inequalities \ref{eq_kinbound}, \ref{eq_b4} and \ref{eq_b3} will be presented in the next section.

\section{Example}\label{sec_example}

% ----------------------- 
% FIGURE 2
\begin{figure*}
	\centering
	\includegraphics[width = 0.85\linewidth]{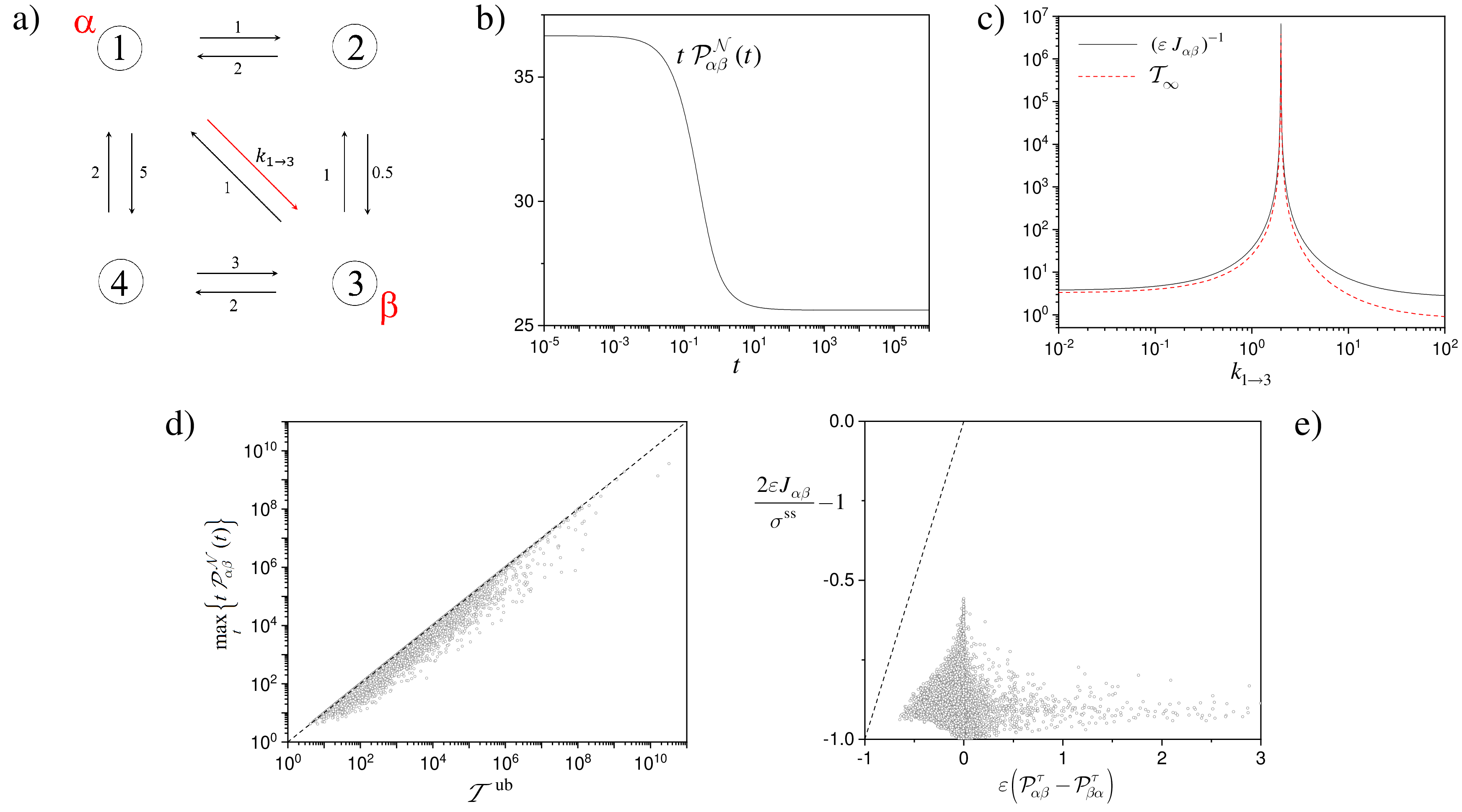} 
\caption{a) The network chosen for the illustrative calculations. All site-site jumps are assumed to occur via single transition channels; the numbers close to the arrows are the values of the corresponding rate constants. b) Temporal profile of $t \, {\cal P}_{\alpha\beta}^{\cal N}(t)$ for $k_{1 \to 3} = 1$. c) Profiles of $(\epsilon J_{\alpha\beta})^{-1}$ and ${\cal T}_{\infty}$ (respectively, the short-time and long-time limits of $t \, {\cal P}_{\alpha\beta}^{\cal N}(t)$) varying $k_{1 \to 3}$. 
d) Illustration of the bound Eq. \ref{eq_kinbound} for $10^4$ randomly generated instances of the network (see the text for details); the dashed line has unit slope. 
In all cases, the maximum value on the ordinate axis was found to be $(\epsilon J_{\alpha\beta})^{-1}$ or ${\cal T}_{\infty}$. 
e) Illustration of the bounds Eq. \ref{eq_b3} (the dashed line has unit slope) and Eq. \ref{eq_b4} (spread on the abscissa, here truncated at the value 3).}
	\label{Fig2}
\end{figure*}
% ----------------------- 

As an example, let us consider the minimal four-site scheme depicted in panel a) of Fig. \ref{Fig2}. We take $\alpha \equiv 1$, $\beta \equiv 3$ and consider the situation in which all site-site jumps occur via single transition channels. Panel b) shows the profile of $t \, {\cal P}_{\alpha\beta}^{\cal N}(t)$ versus $t$ for $k_{1 \to 3} = 1$. In this case the profile is monotonically decreasing and, in agreement with the TUR, it entirely lays (much) above $2 / \sigma^{\rm ss} = 2.81$. Panel c) shows the behavior of the short-time (Eq. \ref{eq_CV_shortt}) and long-time (Eq. \ref{eq_CV_longt}) limits of $t \, {\cal P}_{\alpha\beta}^{\cal N}(t)$ versus $k_{1 \to 3}$ from $10^{-2}$ to $10^2$ keeping all other rate constants fixed. The divergence occurs at a value $k_{1 \to 3}$ for which $J_{\alpha\beta}$ vanishes (although the network is in a nonequilibrium steady state). 

Calculations were then performed for a large number of randomly generated network's instances drawing all rate constants between $10^{-3}$ and $1$ from the uniform distribution on the logarithmic scale. 
Equation \ref{eq_CV_shortt} tells us that the profile of $t \, {\cal P}_{\alpha\beta}^{\cal N}(t)$ versus $t$ initially linearly decreases with slope $r < 0$, while the features at longer times need to be characterized case by case.
In the majority of cases (about $80 \%$) the profiles were monotonically decreasing like the one in panel b); in the other cases the profiles featured an intermediate minimum and a long-time limit ${\cal T}_\infty$ either lower or higher than the short-time value $(\epsilon J_{\alpha\beta})^{-1}$. Other types of more featured profiles were not detected although we cannot exclude their presence for peculiar sets of rate constants. Panel d) illustrates the bound Eq. \ref{eq_kinbound}, while panel e) illustrates the bounds Eq. \ref{eq_b3} (not stringent since the TUR is not either) and Eq. \ref{eq_b4} (look at the values on the abscissa).

\section{Final remarks}\label{sec_conclusions}

In this work we have explored the interrelation between two types of precision coefficients with reference to a transition channel $\alpha \leftrightarrow \beta$ among the many of an irreducible network in which a Markov jump process takes place: precision ${\cal P}_{\alpha\beta}^{\cal N}(t)$ on the integrated current at steady state, and precision on the timing of the $\alpha \to \beta$ (${\cal P}_{\alpha\beta}^{\cal \tau}$) and $\beta \to \alpha$ (${\cal P}_{\beta\alpha}^{\cal \tau}$) recurrences. By resorting to the moment generating function we could derive Eq. \ref{eq_CV}, then used to characterize the precision coefficients and to find interrelations among them. 
In particular, Eqs. \ref{eq_CV_shortt} and \ref{eq_CV_longt} provide the limit forms of ${\cal P}_{\alpha\beta}^{\cal N}(t)$ at short and long time in terms of dynamical observable quantities that implicitly depend on the whole network, but that strictly refer only to $\alpha \leftrightarrow \beta$. In the intermediate timescale, the profile of ${\cal P}_{\alpha\beta}^{\cal N}(t)$ is affected by the time integral of the function $\varphi(t)$ (see Eq. \ref{eq_corr}) which explicitly opens to the rest of the network therefore precluding a transparent interpretation. In addition, inequalities of kinetic (Eqs. \ref{eq_kinbound} and \ref{eq_b4}) and thermodynamic (Eq. \ref{eq_b3}) type could be derived.

The long-time solution (Eq. \ref{eq_CV_longt}) allowed us to retrieve a relation between randomness parameter and Fano factor already known in the context of processive enzymes and molecular motors with irreversible cycle's completion step  \cite{Block95}. On the other hand, the full solution Eq. \ref{eq_corr} extends the interrelation between the different types of precision coefficients to generic networks, finite observation time and $\alpha \leftrightarrow \beta$ reversibility.

Equation \ref{eq_corr} suggests that a deeper and general interrelation should exist between the statistical distributions of ${\cal N}_{\alpha \beta}(t)$ and of the recurrence times $\tau_{\alpha\beta}$, $\tau_{\beta\alpha}$. It could be worth to attempt a formal analysis in this direction going beyond the first two moments of such distributions. Furthermore, a challenge might be to derive the TUR for ${\cal P}_{\alpha\beta}^{\cal N}(t)$ directly from Eq. \ref{eq_CV} or Eq. \ref{eq_corr}, or even get new thermodynamic bounds involving observable features of the $\alpha \leftrightarrow \beta$ channel (including the precision coefficients of the recurrence times) in addition to the global average rate of entropy production $ \sigma^{\rm ss}$. Work on this line is currently in progress.

Finally, we emphasize the novelty and the importance of Eq. \ref{eq_CV} in itself. First, it is useful to perform exact numerical calculations of ${\cal P}_{\alpha\beta}^{\cal N}(t)$. This gives the possibility to explore, for networks of given size, the features of the profile of $t \times {\cal P}_{\alpha\beta}^{\cal N}(t)$ and, especially, to investigate on the conditions (site-site connections and relative values of the jump rate constants) to have a global minimum. Second, Eq. \ref{eq_CV} is potentially a branch point for subsequent elaborations. While in this work we only used Eq. \ref{eq_CV} to arrive at Eq. \ref{eq_corr}, other lines of elaboration might lead to different results if the right-hand side of the equation could be connected to known and relevant quantities of kinetic and thermodynamic type.

\appendix

\section{Derivation of Eq. \ref{eq_CV}}\label{AppA}

The averages $\langle {\cal N}_{\alpha \beta}(t)^n \rangle$ for any integer $n$ can be obtained by exploiting the moment generating function formalism; see for instance ref. \cite{Polettini2019}. In short,
\begin{equation}\label{eq_mom}
\langle {\cal N}_{\alpha \beta}(t)^n \rangle = 
\left. \frac{\partial^n G(q,t)}{\partial q^n} \right|_{q=0}   
\end{equation}
where $G(q,t)$ is the moment generating function given by
\begin{equation}\label{eq_G}
G(q,t) := {\bf 1}^T e^{- t \, {\bf M}(q)} {\bf p}^{\rm ss}    
\end{equation}
where ${\bf 1}^T$ is the row-vector (`T' stands for transpose) with all entries equal to $1$, and
\begin{equation}\label{eq_M}
{\bf M}(q) = {\bf R} + {\bf D}(q)
\end{equation}
in which $\bf R$ is the rate matrix entering the master equation written as $d{\bf p}(t)/dt = - {\bf R} {\bf p}(t)$, that is,
\begin{equation}\label{eq_R}
R_{ij} = -k^{\rm tot}_{j \to i} (1 - \delta_{i,j}) + \delta_{i,j} \sum_{n \neq i} k^{\rm tot}_{i \to n}
\end{equation}
(we recall that $k^{\rm tot}$ stands for the cumulative jump rate constant from one site to the other) and ${\bf D}(q)$ is the $q$-dependent matrix whose elements are
\begin{equation}\label{eq_Ddef}
D_{ij}(q) = \delta_{i,\alpha} \delta_{j,\beta} \, k_{\beta\to\alpha}  \, (1- e^{-q}) 
+ \delta_{i,\beta} \delta_{j,\alpha} \, k_{\alpha\to\beta} \, (1- e^{q}) 
\end{equation}
In particular, up to the second order in $q$ we have
\begin{equation}\label{eq_D}
{\bf D}(q) = q {\bf D}^{(1)} - q^2 {\bf D}^{(2)} + {\cal O}(q^3)
\end{equation}
(here and below, $\cal O$ denotes the order of the remaining terms as $q \to 0$) with the matrices ${\bf D}^{(1)}$ and ${\bf D}^{(2)}$ having elements
\begin{eqnarray}\label{eq_D1D2}
&&D_{ij}^{(1)} = \delta_{i,\alpha} \delta_{j, \beta}  \, k_{\beta\to\alpha}
- \delta_{i,\beta} \delta_{j, \alpha}   k_{\alpha\to\beta}  \nonumber \\
&&D_{ij}^{(2)} = \frac{1}{2} \left(\delta_{i,\alpha} \delta_{j, \beta}  \, k_{\beta\to\alpha} 
+ \delta_{i,\beta} \delta_{j, \alpha}  \,  k_{\alpha\to\beta}  \right) 
\end{eqnarray}

Let us now focus on the power expansion of the matrix exponential in Eq. \ref{eq_G}. All terms of the kind ${\bf R} (\cdots) {\bf R}$ do not contribute since 
${\bf 1}^T {\bf R} = {\bf 0}^T$ (conservation constraint) and ${\bf R} {\bf p}^{\rm ss} = {\bf 0}$ (steady state condition). Thus, by considering Eq. \ref{eq_D}, the only terms that contribute up to $q^2$ are readily identified leading to
\begin{eqnarray}\label{eq_G0}
&&\hspace{-1.0cm}G(q,t) = 1 - t \, q \, {\bf 1}^T {\bf D}^{(1)} {\bf p}^{\rm ss} \cr
&&\hspace{-1.0cm} + q^2 \,  \left[ t \, {\bf 1}^T {\bf D}^{(2)} {\bf p}^{\rm ss} + 
{\bf 1}^T {\bf D}^{(1)} {\bf W}(t) {\bf D}^{(1)} {\bf p}^{\rm ss}
\right]  +  {\cal O}(q^3)
\end{eqnarray}
with
\begin{equation}
{\bf W}(t) = \frac{t^2 }{2} {\bf I} - \frac{t^3}{6}  {\bf R} 
+ \frac{t^4}{24}  {\bf R}^2 
\cdots + \frac{(-t)^n}{n!} {\bf R}^{n-2} + \cdots 
\end{equation}
where $\bf I$ is the identity matrix.

Since $d^2 {\bf W}(t) / d t^2 \equiv e^{-t {\bf R}}$ with ${\bf W}(0)= {\bf 0}$ and 
$d {\bf W}(t)/d t|_{t=0} = {\bf 0}$, it follows that
\begin{equation}
{\bf W}(t) = \int_0^t dt' \int_0^{t'} dt'' e^{-t'' {\bf R}}
\end{equation}
This allows us to write Eq. \ref{eq_G0} as
\begin{eqnarray}\label{eq_Gmom}
G(q,t) &=& 1 + q \, a \, t + q^2 \left[ \frac{b}{2} \, t + \int_0^t dt' \int_0^{t'} dt'' \, g(t'') \right]  \cr
&+& {\cal O}(q^3)
\end{eqnarray}
where $a = - {\bf 1}^T {\bf D}^{(1)} {\bf p}^{\rm ss}$, $b = 2 \, {\bf 1}^T {\bf D}^{(2)} {\bf p}^{\rm ss}$, and $g(t) =  {\bf 1}^T {\bf D}^{(1)} e^{-t {\bf R}} {\bf D}^{(1)} {\bf p}^{\rm ss}$. By using the specific matrix elements given in Eqs. \ref{eq_D1D2} we obtain (recall that $F_{\alpha\beta} =  p_{\alpha}^{\rm ss} \,  k_{\alpha\to\beta}$ and  $F_{\beta\alpha} =  p_{\beta}^{\rm ss} \,  k_{\beta\to\alpha} $)
\begin{equation}\label{eq_ab_def}
a = F_{\alpha\beta} - F_{\beta\alpha} = J_{\alpha\beta}  \;\; , \;\; b = F_{\alpha\beta} + F_{\beta\alpha}
\end{equation}
and
\begin{eqnarray}\label{eq_g0}
&g(t) = - k_{\alpha\to\beta} k_{\beta\to\alpha} \left[ \left( e^{-t {\bf R}}\right)_{\alpha\alpha} \, p_\beta^{\rm ss}  +  \left( e^{-t {\bf R}}\right)_{\beta\beta} \, p_\alpha^{\rm ss}  \right] \cr
&+ k_{\alpha\to\beta}^2 \,  \left( e^{-t {\bf R}}\right)_{\alpha\beta} \, p_\alpha^{\rm ss}
+ k_{\beta\to\alpha}^2 \, \left( e^{-t {\bf R}}\right)_{\beta\alpha} \, p_\beta^{\rm ss}
\end{eqnarray}
By considering that $\left(e^{-t {\bf R}}\right)_{ij} = p(i,t|j)$, Eq. \ref{eq_g0} can be rewritten as
\begin{eqnarray}\label{eq_g1}
g(t) &=& - F_{\alpha\beta} F_{\beta\alpha} \left[ {p(\alpha,t|\alpha)}/{p_\alpha^{\rm ss}}   +  {p(\beta,t|\beta)}/{p_\beta^{\rm ss}}  \right] \cr
&+& F_{\alpha\beta}^2 \,  p(\alpha,t|\beta)/p_\alpha^{\rm ss}
+ F_{\beta\alpha}^2 \,  p(\beta,t|\alpha)/p_\beta^{\rm ss}
\end{eqnarray}
Since $\lim_{t \to \infty} p(i,t|j) = p_i^{\rm ss}$ for any initial $j$, we have that $g_\infty = \lim_{t \to \infty} g(t) = - 2 F_{\alpha\beta} F_{\beta\alpha} + F_{\alpha\beta}^2 + F_{\beta\alpha}^2 = a^2$.
Furthermore, from the definition Eq. \ref{eq_eps} with Eqs. \ref{eq_ab_def} it follows that $\epsilon = a/b$, $F_{\alpha\beta} = (2\epsilon)^{-1}(1+\epsilon) a$ and $F_{\beta\alpha} = (2\epsilon)^{-1}(1-\epsilon) a$. Ultimately, Eq. \ref{eq_g1} takes on
\begin{eqnarray}
g(t) = a^2 -  \frac{a^2}{2} \gamma(t)
\end{eqnarray}
where $\gamma(t)$ is the function in Eq. \ref{eq_gamma}.

By using Eq. \ref{eq_Gmom} in Eq. \ref{eq_mom} we get 
\begin{eqnarray}\label{eq_ab}
&&\langle {\cal N}_{\alpha \beta}(t) \rangle = a \, t \;\; , \nonumber \\
&&\langle {\cal N}_{\alpha \beta}(t)^2 \rangle = b \, t + a^2 \, t^2 
- a^2 \int_0^t dt' \int_0^{t'} dt'' \, \gamma(t'') \cr
&&
\end{eqnarray}
Finally, the form Eq. \ref{eq_CV} for the precision coefficient is obtained by plugging Eqs. \ref{eq_ab} in Eq. \ref{eq_CVdef} and recalling that $a = J_{\alpha\beta} $ and $\epsilon = a/b$.

\section{Derivation of Eq. \ref{eq_rec}}\label{AppB}

Let us consider a generic transition channel $i_1 \to i_2$ and introduce the associated modified rate matrix defined as
\begin{eqnarray}\label{eq_K}
{\bf K} = {\bf R} + {\bf \Delta}
\end{eqnarray}
where $\bf R$ is the rate matrix given in Eq. \ref{eq_R} and $\bf \Delta$ is the matrix with elements
\begin{eqnarray}\label{eq_Delta}
\Delta_{ij} = k_{i_1 \to i_2} \delta_{i,i_2} \delta_{j,i_1}    
\end{eqnarray}
In practice, $\bf K$ is nothing but the original $\bf R$ in which the element on row $i_2$ and column $i_1$ is set to $-k^{\rm tot}_{i_1 \to i_2} + k_{i_1 \to i_2}$ ($0$ in the case of single transition channel). Such a matrix enters the statistics of the survival probabilities conditioned by the clause that $i_1 \to i_2$ has not yet occurred \cite{JCP2019}. In particular, $\rho_{i_1 i_2|s_0}(\tau)=k_{i_1 \to i_2} (e^{- \tau {\bf K}})_{i_1 s_0}$ is the distribution of the first occurrence time $\tau$ of the $i_1 \to i_2$ transition starting from the generic site $s_0$; the distribution of the recurrence time is obtained taking $s_0 = i_2$. The matrix $\bf K$ is invertible and the following relation holds \cite{note2}:
\begin{eqnarray}\label{eq_recK}
\sum_{i} ({\bf K}^{-1})_{i s_0} = \overline{\tau}_{i_1 i_2|s_0}    
\end{eqnarray}
with $\overline{\tau}_{i_1 i_2|s_0}$ the average occurrence time starting from $s_0$.

From the master equation $d {\bf p}(t)/dt = -{\bf R} {\bf p}(t)$ we get 
$-{\bf K} {\bf p} + {\bf \Delta}{\bf p} = d{\bf p}/dt$. The invertibility of $\bf K$ allows us to write
\begin{eqnarray}\label{eq_xxx}
-({\bf I} - {\bf K}^{-1} {\bf \Delta}) {\bf p} = \frac{d}{dt} ({\bf K}^{-1}{\bf p})    
\end{eqnarray}
where $\bf I$ is the identity matrix. Starting from the generic site $s_0$ as initial condition, the $i$-th component of Eq. \ref{eq_xxx} reads
\begin{eqnarray}    
-p(i,t|s_0) + \sum_n ({\bf K}^{-1} {\bf \Delta})_{in} p(n,t|s_0) = \cr
= \frac{d}{dt} \sum_n ({\bf K}^{-1})_{in} p(n,t|s_0)
\end{eqnarray}
By recalling Eq. \ref{eq_Delta}, the summation on the left-hand side simplifies leading to
\begin{eqnarray}    
-p(i,t|s_0) + k_{i_1 \to i_2} ({\bf K}^{-1})_{i i_2} p(i_1,t|s_0) = \cr
= \frac{d}{dt} \sum_n ({\bf K}^{-1})_{in} p(n,t|s_0)
\end{eqnarray}
Let us now take the summation over $i$ at both members and make use of Eq. \ref{eq_recK} also considering that $\overline{\tau}_{i_1 i_2|i_2} \equiv \overline{\tau}_{i_1 i_2} = (p_{i_1}^{\rm ss} k_{i_1 \to i_2})^{-1}$. This yields
$-1 + p(i_1,t|s_0)/p_{i_1}^{\rm ss} = (d/dt)(\sum_n \overline{\tau}_{i_1 i_2|n} p(n,t|s_0))$, where the left-hand side corresponds exactly to $\chi_{i_1 s_0}(t)$. Thus,
\begin{eqnarray}\label{eq_rec0}    
\chi_{i_1 s_0}(t)= 
\frac{d}{dt} \sum_n \overline{\tau}_{i_1 i_2|n} p(n,t|s_0)
\end{eqnarray}
Let us note that Eq. \ref{eq_rec0} holds for any choice of $i_2 \neq i_1$ on condition that $i_2$ be directly reachable from $i_1$, and for any choice of the transition channel connecting $i_1$ to $i_2$ (in the case of multiple channels).

The time integration of Eq. \ref{eq_rec0} finally yields Eq. \ref{eq_rec} where $i_1$ and $i_2$ are replaced by $i$ and $j$ directly connected by $i \to j$.

\section{Numerical solution of Eq. \ref{eq_CV} for diagonalizable rate matrices}\label{AppC}

The conditional probability that enters Eq. \ref{eq_chi} corresponds to $p(i,t|s_0) = (e^{-t {\bf R}})_{i s_0}$ where $\bf R$ is the rate matrix of the master equation (see Eq. \ref{eq_R}). Let us consider the most typical case in which $\bf R$ is diagonalizable, i.e., all eigenvalues are distinct or, in the case of degeneracies, a complete set of independent eigenvectors can be however determined (the peculiar case of non-diagonalizable matrix \cite{note_no_diag_R} requires a bit more complex elaboration; see for instance ref. \cite{net1976}). In this case, the matrix exponential is handled by diagonalizing $\bf R$. This leads to
$p(i,t|s_0) = ({\bf V} e^{-t {\bf \Lambda}} {\bf V}^{-1})_{i s_0}$ where $\bf V$ is the matrix whose columns are the right-eigenvectors of $\bf R$, and $\bf \Lambda$ is the diagonal matrix of the eigenvalues. In nonequilibrium conditions, the eigenvalues
are generally complex, $\lambda_n = \lambda_n^{\rm R} + \imath \lambda_n^{\rm I}$, pair-conjugated and with real parts strictly positive except for a unique null eigenvalue (associated to the steady state distribution) which does not contribute to $\chi_{i s_0}(t)$. Explicitly,
$\chi_{i s_0}(t) = \sum_{n \neq n_0} w_{i s_0}(n) e^{-\lambda_n t}$ where $\lambda_{n_0} = 0$ is the null eigenvalue and $w_{i s_0}(n) = V_{in} ({\bf V}^{-1})_{n s_0} / p_i^{\rm ss}$ in which the steady-state probabilities $p_i^{\rm ss}$ correspond to the elements $V_{i n_0}$ normalized to have sum one. 

The double time-integral in Eq. \ref{eq_CV} is analytical taking into account that for each of the four contributions we have
\begin{eqnarray}
&&\frac{1}{t^2}\int_0^t dt' \int_0^{t'} dt'' \chi_{i s_0}(t'') =  \sum_{n \neq n_0} 
w_{i s_0}(n)  f_n(t) \;\; ,\cr
&&f_n(t)=\frac{1}{\lambda_n t} - \frac{1-e^{-\lambda_n t}}{(\lambda_n t)^2} 
\end{eqnarray}
Note that $f_n(t) \to 1/2$ as $t \to 0$, while $f_n(t) \simeq (\lambda_n t)^{-1}$ in the long time limit. Thus, at times much longer than $(\min_{n \neq n_0} \{\lambda_n^{\rm R} \})^{-1}$ we have that ${\cal P}_{\alpha\beta}^{\cal N}(t) \propto t^{-1}$ as at short times (see Eq. \ref{eq_CV_shortt}), but with a lower or higher proportionality coefficient.

\section{Derivation of Eq. \ref{eq_corr}}\label{AppD}

By multiplying both members of Eq. \ref{eq_CV} by $t^2$ and taking the time derivative, we get
\begin{equation}\label{eq_start}
\frac{d}{dt}[t^2 {\cal P}_{\alpha\beta}^{\cal N}(t)]  = \frac{1}{\epsilon J_{\alpha\beta}} - \int_0^t dt' \, \gamma (t')  
\end{equation}
The time integral on the right-hand side is obtained from Eq. \ref{eq_gamma} by elaborating the four single integrals of $\chi_{\alpha\alpha}(t')$, $\chi_{\alpha\beta}(t')$, $\chi_{\beta\alpha}(t')$ and $\chi_{\beta\beta}(t')$. Equation \ref{eq_rec2} is now employed making, in each case, the specific assignment of $i$, $s_0$ and $j$ to elaborate such integrals. Specifically, 
in the bidirectional case all four integrals contribute and
we set $i=\alpha$, $s_0 = \alpha$, $j = \beta$ for $\chi_{\alpha\alpha}$, $i=\beta$, $s_0 = \beta$, $j = \alpha$ for $\chi_{\beta\beta}$, $i=\alpha$, $s_0 = \beta$, $j = \beta$ for $\chi_{\alpha\beta}$, and $i=\beta$, $s_0 = \alpha$, $j = \alpha$ for $\chi_{\beta\alpha}$. For one-directional $\alpha \to \beta$, only the contribution of $\chi_{\alpha\beta}$ survives since $c_0 = c_- = 0$ while $c_+ = 2$.

Let us give some relations which will be of use later. The following ones are readily derived from the definitions in Eq. \ref{eq_cs}: 
\begin{eqnarray}\label{eq_w1}
c_0 + c_\pm = \frac{1\pm \epsilon}{\epsilon^2}
\end{eqnarray}
and
\begin{eqnarray}\label{eq_w1b}
c_0 - c_+ = -\frac{1 + \epsilon}{\epsilon} \;\; , \;\; 
c_0 - c_- = \frac{1 - \epsilon}{\epsilon} 
\end{eqnarray}
The following relations are obtained from the definition of $\epsilon$ given in Eq. \ref{eq_eps} and from the relations in Eq. \ref{taus} between average recurrence times and fluxes: 
\begin{eqnarray}\label{eq_w2}
\frac{1+\epsilon}{\epsilon} = \frac{2}{\overline{\tau}_{\alpha \beta} \, J_{\alpha\beta}}  \;\; , \;\; 
\frac{1-\epsilon}{\epsilon} = \frac{2}{\overline{\tau}_{\beta\alpha} \, J_{\alpha\beta}} 
\end{eqnarray}
and
\begin{eqnarray}\label{eq_w3}
\frac{c_+}{c_0} = \frac{\overline{\tau}_{\beta\alpha}}{\overline{\tau}_{\alpha\beta}} \;\; , \;\;
\frac{c_-}{c_0} = \frac{\overline{\tau}_{\alpha\beta}}{\overline{\tau}_{\beta\alpha}}
\end{eqnarray}
It is implicit that here we deal with the unbalanced case $\epsilon \neq 0$, and that the above relations have to be understood in the one-directional limit cases.

Let us first consider the general bidirectional case $\epsilon \neq \pm 1$. With the assignments of $i$, $s_0$, $j$ given above, from Eq. \ref{eq_rec2} we get
\begin{eqnarray}\label{eq_aa}
\int_0^t dt' \, \chi_{\alpha\alpha} (t') &=& 
-\overline{\tau}_{\alpha\beta|\alpha} 
+ \frac{\overline{\tau}_{\alpha\beta}}{2} + \frac{\overline{\tau}_{\alpha\beta}}{2} {\cal P}^\tau_{\alpha \beta}  \cr
&&+ \sum_{n} \overline{\tau}_{\alpha\beta|n} \, p_n^{\rm ss} \, \chi_{n \alpha}(t) \;\; , \cr
\int_0^t dt' \, \chi_{\beta\beta} (t') &=& 
-\overline{\tau}_{\beta\alpha|\beta} 
+ \frac{\overline{\tau}_{\beta\alpha}}{2} 
+ \frac{\overline{\tau}_{\beta\alpha}}{2} {\cal P}^\tau_{\beta \alpha}  \cr
&&+ \sum_{n} \overline{\tau}_{\beta\alpha|n} \, p_n^{\rm ss} \, \chi_{n \beta}(t) 
\;\; , \cr
\int_0^t dt' \, \chi_{\alpha\beta} (t') &=& 
-\frac{\overline{\tau}_{\alpha\beta}}{2} 
+\frac{\overline{\tau}_{\alpha\beta}}{2} {\cal P}^\tau_{\alpha \beta}  \cr
&&+ \sum_{n} \overline{\tau}_{\alpha\beta|n} \, p_n^{\rm ss} \, \chi_{n \beta}(t) 
\;\; , \cr
\int_0^t dt' \, \chi_{\beta\alpha} (t') &=& 
-\frac{\overline{\tau}_{\beta\alpha}}{2} 
+ \frac{\overline{\tau}_{\beta\alpha}}{2} {\cal P}^\tau_{\beta\alpha}  \cr
&&+ \sum_{n} \overline{\tau}_{\beta\alpha|n} \, p_n^{\rm ss} \, \chi_{n \alpha}(t)
\end{eqnarray}
where it has been made use of $\overline{\tau}_{\alpha\beta|\beta} \equiv \overline{\tau}_{\alpha\beta}$ and
$\overline{\tau}_{\beta\alpha|\alpha} \equiv \overline{\tau}_{\beta\alpha}$. 
Plugging Eqs. \ref{eq_aa} into the time-integrated form of Eq. \ref{eq_gamma}, we get
\begin{eqnarray}\label{eq_gamma_bid}
\int_0^t dt' \, \gamma (t') = A_1 + A_2 + A_3 + A_4(t)    
\end{eqnarray}
where the various $A$ on the right-hand side are addends that derive from a suitable grouping of the terms. Specifically,
\begin{eqnarray}\label{eq_As} 
&&A_1 = \overline{\tau}_{\alpha\beta} (c_0 + c_+)/2 + \overline{\tau}_{\beta\alpha} (c_0 + c_-)/2 \;\;, \nonumber \\
&&A_2 = -c_0 \, (\overline{\tau}_{\alpha\beta|\alpha} + \overline{\tau}_{\beta\alpha|\beta}) 
\;\;, \nonumber \\
&&A_3 = {\cal P}^\tau_{\alpha\beta} \, \overline{\tau}_{\alpha\beta} (c_0 - c_+)/2 +
{\cal P}^\tau_{\beta\alpha} \, \overline{\tau}_{\beta\alpha} (c_0 - c_-)/2 \;\;, \nonumber \\
&&A_4(t) = \sum_n [
c_0 \, \overline{\tau}_{\alpha\beta|n} \, \chi_{n\alpha}(t)
+ c_0 \, \overline{\tau}_{\beta\alpha|n} \, \chi_{n\beta}(t) \cr
&&\hspace*{1cm} - c_+ \, \overline{\tau}_{\alpha\beta|n} \, \chi_{n\beta}(t)
- c_- \, \overline{\tau}_{\beta\alpha|n} \, \chi_{n\alpha}(t)
] \, p_n^{\rm ss}
\end{eqnarray}
By inserting Eq. \ref{eq_gamma_bid} into Eq. \ref{eq_start} and making the time integration, we obtain a form of ${\cal P}_{\alpha\beta}^{\cal N}(t)$ akin to Eq. \ref{eq_corr} with
\begin{eqnarray}\label{eq_elab}
{\cal T}_\infty = \frac{1}{\epsilon J_{\alpha\beta}}- A_1 - A_2 - A_3 \;\; , \;\;
\varphi(t) = - A_4(t)
\end{eqnarray}
By using the relations in Eq. \ref{eq_w3} in combination with those in Eqs. \ref{eq_w1} and \ref{eq_w2}, $A_1$ boils down to
\begin{eqnarray}\label{eq_A1}   
A_1 = \frac{2}{\epsilon \, J_{\alpha\beta}}
\end{eqnarray}
The addend $A_2$ is already in its final form. By using the relations in Eqs. \ref{eq_w1b} and \ref{eq_w2}, $A_3$ becomes
\begin{eqnarray}\label{eq_A3}   
A_3 = ({\cal P}^\tau_{\beta\alpha} - {\cal P}^\tau_{\alpha\beta})/J_{\alpha\beta}
\end{eqnarray}
Finally, by factoring out $c_0$ and then employing Eqs. \ref{eq_w3}, the expression of $A_4(t)$ given in Eq. \ref{eq_As} becomes
\begin{eqnarray}\label{eq_A4}   
A_4(t) &=& c_0 \, \sum_n [
\overline{\tau}_{\alpha\beta|n} \, \chi_{n\alpha}(t)
+ \overline{\tau}_{\beta\alpha|n} \, \chi_{n\beta}(t) \cr
&&  -\frac{\overline{\tau}_{\beta\alpha}}{\overline{\tau}_{\alpha\beta}}  \, \overline{\tau}_{\alpha\beta|n} \, \chi_{n\beta}(t)
-\frac{\overline{\tau}_{\alpha\beta}}{\overline{\tau}_{\beta\alpha}}  \, \overline{\tau}_{\beta\alpha|n} \, \chi_{n\alpha}(t)
] \, p_n^{\rm ss} \nonumber \\
&\equiv& c_0 \, \sum_n \left( 
\frac{\overline{\tau}_{\alpha\beta|n}}{\overline{\tau}_{\alpha\beta}}
- \frac{\overline{\tau}_{\beta\alpha|n}}{\overline{\tau}_{\beta\alpha}}
\right) \times \cr
&&\hspace*{1cm}\times ( \overline{\tau}_{\alpha\beta} \chi_{n\alpha}(t) - 
\overline{\tau}_{\beta\alpha} \chi_{n\beta}(t) ) \, p_n^{\rm ss} \cr
&&
\end{eqnarray}
The use of these forms of $A_1$, $A_2$, $A_3$ and $A_4(t)$ in Eq. \ref{eq_elab} yields the expressions of ${\cal T}_\infty$ and $\varphi(t)$ given in Eqs. \ref{tauinft_bid} and \ref{eq_phi_bid} of the main text for the bidirectional case.

In the one-directional case $\alpha \to \beta$, we have that $\gamma(t) = -c_+ \chi_{\alpha\beta}(t) = - 2 \chi_{\alpha\beta}(t)$. Thus, 
\begin{eqnarray}\label{eq_gamma_one}
\int_0^t dt' \, \gamma (t') =\overline{\tau}_{\alpha\beta} 
-{\cal P}^\tau_{\alpha\beta}\overline{\tau}_{\alpha\beta} + B(t)    
\end{eqnarray}
where
\begin{eqnarray}\label{eq_Bs}
B(t) = -2 \sum_n \overline{\tau}_{\alpha\beta|n} \, \chi_{n \beta}(t) \, p_n^{\rm ss}
\end{eqnarray}
Let us note that, for the present case $\epsilon = 1$, we have $J_{\alpha\beta} = F_{\alpha\beta} = \overline{\tau}_{\alpha\beta}^{-1}$, hence the first addend $(\epsilon J_{\alpha\beta})^{-1}$ in Eq. \ref{eq_start} becomes $\overline{\tau}_{\alpha\beta}$. Thus, the use of Eq. \ref{eq_gamma_one} in Eq. \ref{eq_start} eventually leads to a relation akin to Eq. \ref{eq_corr} with assignments
\begin{eqnarray}\label{eq_elab2}
{\cal T}_\infty = {\cal P}^\tau_{\alpha\beta} \overline{\tau}_{\alpha\beta} \;\; , \;\; \varphi(t) = - B(t)
\end{eqnarray}
corresponding to Eqs. \ref{tauinft_oned} and \ref{eq_phi_oned} of the main text.

\section*{Acknowledgments}
The authors acknowledge the financial contribution from ``Fondazione Cassa di Risparmio di Padova e Rovigo'' (CARIPARO) within the framework of the project ``NoneQ'', ID 68058.

\bibliographystyle{unsrt}

\begin{thebibliography}{10}
\bibitem{note1}
The total $\sigma^{\rm ss}$ is obtained by adding the contributions $(F_{ij} - F_{ji})\ln (F_{ij}/F_{ji})$ over all transition channels of the $i \leftrightarrow j$ pair, and then adding over all pairs of connected sites. To be adherent to the degree of resolution at disposal, $\sigma^{\rm ss}$ should take into account all visible transition channels, even because a coarser description (lumping) would produce a smaller value of $\sigma^{\rm ss}$ and, consequently, a less tight TUR.
\bibitem{note_gamma0}
Specifically, from Eqs. \ref{eq_gamma}, \ref{eq_chi} and \ref{eq_cs} we get $\gamma(0) = 2 + c_0 (1/{p_\alpha^{\rm ss}} + 1/{p_\beta^{\rm ss}})$.
\bibitem{note3b}
The numerical route was also used to check the correctness of Eq. \ref{eq_CV} by comparing the results with the values from stochastic simulations using Gillespie's algorithm. The check was done for the four-site network.
\bibitem{note_bound}
The starting point is to consider the four contributions to $\gamma(t'')$ in Eq. \ref{eq_gamma}, and then apply Eqs. \ref{eq_bxi1} and \ref{eq_bxi2} in which the sites $i$ and $j$ are $\alpha$ and $\beta$, or $\beta$ and $\alpha$. This leads to $\int_0^{t'} dt'' \gamma(t'') \geq -(c_+ \overline{\tau}^{\rm max}_{\alpha\beta} + c_- \overline{\tau}^{\rm max}_{\beta\alpha}) + c_+ \overline{\tau}_{\alpha\beta} + c_- \overline{\tau}_{\beta\alpha}$. The bound in Eq. \ref{eq_kinbound} with Eq. \ref{eq_b0} is finally obtained by considering that $c_+ \overline{\tau}_{\alpha\beta} + c_- \overline{\tau}_{\beta\alpha} = 2/(\epsilon J_{\alpha\beta})$ and performing the integration on $t'$ in Eq. \ref{eq_CV}. Let us stress that Eqs. \ref{eq_bxi1} and \ref{eq_bxi2} are applicable only if the transition $i \to j$ is feasible, while the transition channel $\alpha \leftrightarrow \beta$ might be one-directional. However, in case of irreversibility, $c_+$ or $c_-$ would be zero in Eq. \ref{eq_gamma} thereby removing the critical terms
  from beginning and ensuring the correct final result.
\bibitem{note5}
Eq. \ref{tauinft_bid} with $\overline{\tau}_{\alpha\beta|\alpha} < \overline{\tau}_{\alpha\beta}$ and $\overline{\tau}_{\beta\alpha|\beta} < \overline{\tau}_{\beta\alpha}$ yields $\epsilon \, ({\cal P}_{\alpha\beta}^\tau - {\cal P}_{\beta\alpha}^\tau) > 2 \epsilon J_{\alpha\beta}/\sigma^{\rm ss} + 1 - \epsilon J_{\alpha\beta} c_0 (\overline{\tau}_{\alpha\beta} + \overline{\tau}_{\beta\alpha})$. Eq. \ref{eq_b3} is obtained by considering that $\epsilon J_{\alpha\beta} c_0 (\overline{\tau}_{\alpha\beta} + \overline{\tau}_{\beta\alpha}) = 2$.
\bibitem{note2}
The average occurrence time is given by $\overline{\tau}_{i_1 i_2 | s_0} = \int_0^\infty d\tau \, \tau \rho_{i_1 \to i_2 | s_0}(\tau) = k_{i_1 \to i_2} ({\bf K}^{-2})_{i_1 s_0} = k_{i_1 \to i_2} \sum_n ({\bf K}^{-1})_{i_1 n} ({\bf K}^{-1})_{n s_0}$. The normalization $\int_0^\infty d\tau \rho_{i_1 \to i_2 | n}(\tau) = k_{i_1 \to i_2} ({\bf K}^{-1})_{i_1 n} = 1$ (from which we get $({\bf K}^{-1})_{i_1 n} = 1/ k_{i_1 \to i_2}$ for any $n$) leads to Eq. \ref{eq_recK}.
\bibitem{note_no_diag_R}
For example, a four-site irreducible network with non-diagonalizable rate matrix is the one with connections $1 \rightleftharpoons 2$, $2 \to 3$, $3 \to 4$, $4 \to 2$ and all jump rate constants having the same value.
\end{thebibliography}

\end{document}